\documentclass [aps,prl,superscriptaddress,showpacs,twocolumn,letterpaper] {revtex4}

\usepackage {graphicx}
\DeclareGraphicsExtensions {.eps}

\newcommand {\bisco} {Bi$_2$Sr$_2$CaCu$_2$O$_{8+\delta}$}
\newcommand {\sixteeno} {$^{16}$O}
\newcommand {\sixteenor} {$^{16}$O$_R$}
\newcommand {\eighteeno} {$^{18}$O}

\begin{document}

\title {Strong and complex electron-lattice correlation in optimally doped \bisco}
\author {G.-H. Gweon}
\affiliation {Department of Physics, University of California, Berkeley,
  CA, 94720}
\affiliation {Department of Physics, University of California, Santa Cruz,
  CA, 95064}
\author {S. Y. Zhou}
\author {M. C. Watson}
\affiliation {Department of Physics, University of California, Berkeley,
  CA, 94720}
\author {T. Sasagawa}
\affiliation {Department of Advanced Materials Science, University of
  Tokyo, Kashiwa, Chiba 277-8561, Japan}
\affiliation {CREST, Japan Science and Technology Agency, Saitama
  332-0012, Japan}
\author {H. Takagi}
\affiliation {Department of Advanced Materials Science, University of
  Tokyo, Kashiwa, Chiba 277-8561, Japan}
\affiliation {CREST, Japan Science and Technology Agency, Saitama
  332-0012, Japan}
\affiliation {RIKEN (The Institute of Physical and Chemical Research),
  Wako 351-0198, Japan}
\author {A. Lanzara}
\affiliation {Department of Physics, University of California, Berkeley,
  CA, 94720}
\affiliation {Materials Sciences Division, Lawrence Berkeley National Laboratory, Berkeley, CA, 94720}
\date {\today}

\begin {abstract}

We discuss the nature of electron-lattice interaction in optimally doped \bisco~samples, using isotope effect (IE) in angle resolved photoemission spectroscopy (ARPES) data.
The IE in the ARPES line width and the IE in the ARPES dispersion are both quite large, implying a strong electron-lattice correlation.
The strength of the electron-lattice interaction is ``intermediate,'' i.e.~stronger than the Migdal-Eliashberg regime but weaker than the small polaron regime, requiring a more general picture of the ARPES ``kink'' than the commonly used Migdal-Eliashberg picture.
The two IEs also imply a complex interaction, due to their strong momentum dependence and their differing sign behaviors.
In sum, we propose an intermediate-strength coupling of electrons to localized lattice vibrations via charge density fluctuations.

\end {abstract}



\maketitle

In the last few years, several angle resolved photoemission spectroscopy (ARPES) reports suggested a significant interaction of dynamic lattice distortions with electrons in the cuprate superconductors \cite {lanzara, gweon, gweon-nature, cuk}, but the basic nature of the electron-lattice interaction is still unclear.

On one hand, the Migdal-Eliashberg theory, a standard model of the electron-phonon interaction in solids, seems to provide a basic framework for explaining \cite {verga,devereaux} a key feature of ARPES data, i.e.~the ``kink'' in the ARPES dispersion \cite {lanzara, gweon, cuk}.  In this theory, the ARPES kink appears at phonon energy, $\omega_p$.  For excitation energy ($\omega$) smaller than $\omega_p$, long-lived quasi-particles (electron dressed by virtual phonons) form.  For $\omega > \omega_p$, short-lived electrons scatter strongly with real phonons.  For $\omega \gg \omega_p$, the effect of the scattering becomes negligible.  This familiar Migdal-Eliashberg picture has been also generalized to cases where other bosonic excitations are involved \cite {grilli,norman}.

On the other hand, a large number of experimental reports \cite {calvani, ruani, heeger, bianconi, bozovic, billinge, mihailovic, imai, kochelaev, gweon-nature, shen} suggest an interaction strength beyond the Migal-Eliashberg theory.  Such a proposal is plausible also from the ARPES point of view, because as momentum approaches the Brillouin zone boundary, $\omega_p$ ($\approx 70$ meV) becomes comparable to the effective band width governed by Fermi velocity \cite {perali}, weakening the Migal-Eliashberg theory, and the actual quasi-particle weight $Z$ measured by ARPES is rather small, $\sim$ 0.1 \cite {feng}.

To date, there has been no ARPES study reconciling these two important points of view.  This absence appears to be due to the complexity of the cuprate physics, which makes it difficult to disentangle the effect of electron-lattice interaction from the effect of strong electron-electron interaction.  As we demonstrated recently \cite {gweon-nature}, ARPES on isotope substituted samples is a unique method that can directly sort out the role of the lattice to shed light on this complexity.

In this Letter, we provide evidence of an anomalously large isotope effect (IE) in the ARPES peak width for optimally doped \bisco~superconductor, in addition to the IE in the ARPES peak position reported earlier \cite {gweon-nature}.  Detailed comparison of the data with various theories of the electron-lattice interaction provides compelling evidence that the strength of the interaction is intermediate, i.e.~beyond the Migal-Eliashberg regime but not in the small polaron regime.  This and another feature, the local nature of the lattice vibrations involving only a few lattice sites \cite {bianconi-PRL, grilli}, are an important refinement of an anisotropic electron-lattice interaction model proposed before \cite {gweon, cuk, devereaux}.  Finally, we believe that our results give important clues for understanding the nodal-anti-nodal dichotomy observed in the cuprates \cite {xingjiang, shen} as well as the interplay between the electron-electron interaction and the electron-lattice interaction, as proposed recently \cite {gweon-nature, seidel, fu}.

We discuss data taken at low temperature (25 K), where the electron-lattice interaction is strongly enhanced \cite {gweon-nature} well below $T^*$ (pseudo-gap temperature) $\approx T_c$ (92 K for \sixteeno~and 91 K for \eighteeno).
Details of the experiment are described elsewhere \cite {gweon-nature}.
Throughout this Letter, we use the term ``high energy'' or ``low
energy'' to refer to the magnitude $|E - E_F|$ ($E_F$ is the
Fermi energy), high or low meaning relative to the kink energy in particular.

\begin {figure} [!t]

\includegraphics [width = 3.0 in] {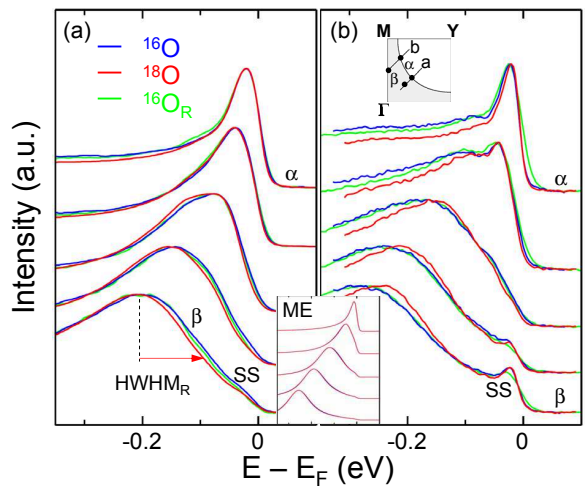}

\caption {Comparison of ARPES EDCs for three samples with different isotope treatments, normalized to same peak height; normal (\sixteeno), isotope-substituted (\eighteeno) and isotope re-substituted (\sixteenor) samples for cuts a and b, as indicated in the inset of panel b along with the Fermi surface.  Data for each angle were integrated over a 1$^\circ$ angle window in favor of statistics.  The inset saddling the two panels shows Migal-Eliashberg simulations for \sixteeno~and \eighteeno, as described in text.}

\end {figure}


In Figure 1 we show raw ARPES data along two important momentum space cuts, cut a (nodal) and cut b (off-nodal or near-anti-nodal \cite {full-momentum-dep-later}).  A more detailed momentum dependence is described elsewhere \cite {gweon-nature, shift} for the peak position and below for the width \cite {full-momentum-dep-later}.
The energy distribution curves (EDCs; intensity as function of energy at fixed momentum value) displayed here show clear IE for both the peak position and the peak width.
Previously \cite {gweon-nature}, momentum distribution curves (MDCs; intensity as function of momentum at fixed energy value) were used to discuss IE in the MDC dispersion and a small red-shift of the kink energy.
In this Letter, we go further to discuss the line width, which requires more care in analysis.
For this reason, we discuss EDCs rather than MDCs and use the half width at half maximum on the right side (HWHM$_R$) of EDC peak to quantify the peak width \cite {explain-HWHM-R}, as illustrated in panel a.
As we will show, the EDC analysis gives the same result as the MDC analysis in terms of the peak position.

Two observations can be added \cite {gweon-nature} to support the intrinsic nature of the observed IE\@.  
First, high energy peaks in Figure 1 {\em sharpen up} upon isotope substitution, reinforcing the argument \cite
{gweon-nature} that a simple band structure effect or a simple disorder
effect, arising from a stronger disorder for the \eighteeno~sample due to
the substitution process, cannot be the origin of the observed IE\@. 
Second, the EDCs for the re-substituted sample (green) are nearly identical to those of the \sixteeno~sample.

It is quite clear that the Migal-Eliashberg theory (inset) predicts a small IE overall, in strong contrast to the data.
For the Migal-Eliashberg theory, we chose the finite temperature formalism \cite {verga}
for electron coupled to seven Einstein phonon modes, equally spaced from 10
to 70 meV, with the maximum determined by neutron scattering data \cite {egami}.  
For this model, we refer to this maximum frequency as $\omega_{p}$.
For the purpose of this paper, this choice of phonon frequencies is equivalent to a more realistic \cite {egami} choice of a fewer number of high frequency modes, while doing a better job of describing a smooth kink \cite {gweon-nature} seen in the data.
We use a gap-less linear band with $v_F$ $=1.8$ eV\AA~for $\epsilon (k)$, and 0.5 for the dimensionless
electron-phonon coupling constant $\lambda$ \cite {verga}.
For \eighteeno, only $\omega_p$ is modified, in accordance with the harmonic phonon model.
We have checked extensively that use of additional features such as anisotropic coupling \cite {devereaux-private-comm}, full band structure and superconducting gap \cite {grilli, norman, devereaux} does not affect the discussion presented below.

\begin {figure} [!b]

\includegraphics [width = 3.0 in]{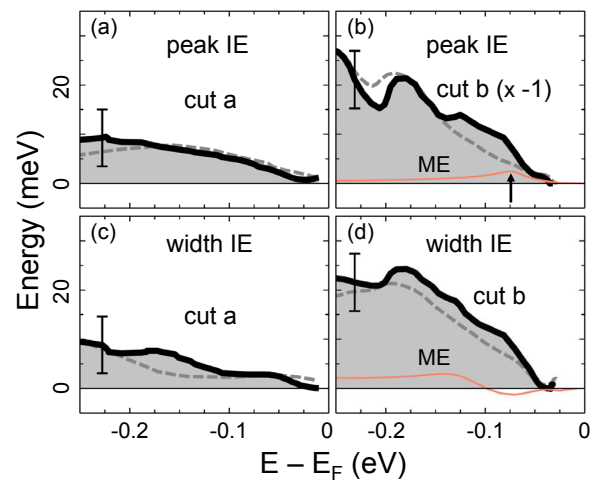}

\caption {(a,b) IE in the EDC peak positions for cuts a and b.  (c,d) IE in the EDC widths (HWHM$_R$).  The IE is defined as \sixteeno$-$\eighteeno, and note that the sign-reversed IE is plotted in panel b.  The horizontal energy axis correponds to the peak position for the \sixteeno~sample.  Gray dashed lines are IEs for \sixteenor.  For comparison Migal-Eliashberg simulation results are shown as light red lines in panels b,d.}

\end {figure}

In Figure 2, we present a detailed analysis of the IE introduced in Figure 1.
Panels a and b show ``peak IE,'' namely the isotope dependence of EDC peak position.
Likewise, panels c and d show ``width IE,'' the isotope dependence of EDC peak width.
Several important points can be made.
First, both IEs are large at high energy, while small at low energy \cite {note-T-dep}.
This result shows a complete agreement with the previous work \cite {gweon-nature} as the peak IE is concerned, and further shows that the kink energy ($\approx$ the arrow position in panel b for both cuts) continues to separate small IE and large IE also for the width IE, as it separates the sharp coherent peak and the broad incoherent peak \cite {gweon-nature}.
Second, the Migal-Eliashberg simulation (light red lines in b,d) poorly explains the data in general and the data at high energy in particular.
Third, the IEs are enhanced for the off-nodal cut, consistent with a strongly momentum dependent electron-lattice coupling \cite {devereaux} and making the failure of the Migal-Eliashberg theory less severe for the nodal cut.
Fourth, the sign of the width IE is unchanged going from cut a to cut b, i.e.~the peak always sharpens up for \eighteeno, while the sign of the peak IE changes from cut a to cut b.
These differing sign behaviors are difficult to understand within the usual self energy analysis scheme \cite {fink}, because the Kramers-Kronig relationship connects the real part and the imaginary part of self energy \cite {fink}, determined mainly by the peak position and width respectively.
This means that the self energy analysis scheme as it has been employed to date under two key assumptions, momentum independent self energy and high energy data cutoff at $\approx 0.3-0.4$ eV, is unreliable to the accuracy relevant here.

\begin {figure} [!t]

\includegraphics [width = 3.0 in]{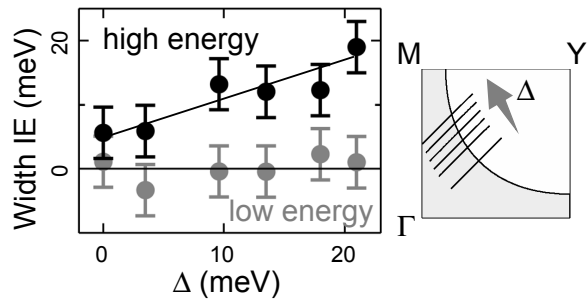}

\caption {
  IE of the EDC width as a function of the superconducting gap $\Delta$,
  for six momentum cuts \cite {full-momentum-dep-later} 
  shown in the right panel.  For each cut, low
  energy (-70 meV to 0 meV) average and high energy (-250 meV to
  -70 meV) average of IE are shown with error bars. The line shown for high energy
  IE is guide to the eye.}

\end {figure}

In Figure 3, we summarize the momentum dependence of the width IE up to cut b \cite {full-momentum-dep-later}.
The low energy IE is small, and vanishes within error bars.  
In contrast, the high energy IE is finite and shows an approximately linear correlation with the superconducting gap up to cut b \cite {full-momentum-dep-later}, as the peak IE \cite {gweon-nature}. 
The corresponding IEs for the Migal-Eliashberg theory is negligible, 2 meV at the most.

The failure of the Migal-Eliashberg theory can be traced back to the the single phonon loop approximation for electron self energy, subsequently resulting in a small IE (shown by light red lines in Figures 2b and 2d) governed at the most by the isotope dependence of the single phonon energy.  For instance, within this theory the high energy line width is given by $\approx \omega_p$, explaining the small IE at high energy (Figure 2d) as well as explaining why the ARPES width of this theory is by a factor of $\approx 2$ too small in comparison to the observed width (Figure 1).
In the literature, there have been two different approaches to remedy this failure of the Migal-Eliashberg theory.  First, Seibold and Grilli \cite {grilli} used a more general Migal-Eliashberg model which involves $\lambda$ that is strongly momentum and {\em isotope} dependent.  The physical reason, especially for the latter, is that the boson that couples to electrons in this case is critical charge order fluctuations of correlation length of a few lattice constants, instead of phonons extended throughout the lattice.  Using this generalization, it was shown that the model can qualitatively explain the sign change seen in Figures 2a,b but not the IE seen in Figures 2c,d, predicting a broader peak for \eighteeno.
Second, models \cite {mishchenko, fratini} incorporating interaction strength going beyond the Migal-Eliashberg theory were shown to be able to explain the large (small) isotope effect at high (low) energy and also the {\em sharpening} of ARPES peak induced by \eighteeno.
On the other hand, the sign change of Figures 2a,b could not be addressed, presumably due to the simple form of interaction assumed.
It is significant to note that in these calculations the binding energy of peak increases while the peak width decreases, a non-trivial result in qualitative agreement with our nodal cut data (Figures 2a,c).

Based on these results, we propose that a combination of these two approaches is required to be totally successful, as we further elaborate now.


\begin {figure}[!b]

\includegraphics [width = 3.2 in]{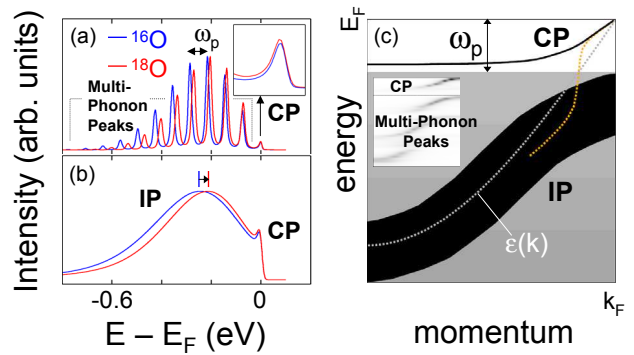}

\caption {
  (a,b) Simulation of EDCs at $k = k_F$ for cut b using
  a small polaron theory \cite {alexandrov}. The line shape simulations
  include a 10 meV (panel a) and an additional 100 meV (panel b) FWHM Lorentzian broadening, 
  the latter only for multi-phonon peaks.
  (c) Proposed schematic diagram of ARPES kink for intermediate and weak electron-lattice interaction.
  The thickness of the line used for CP or IP dispersion crudely represents the ARPES line width.
  The intensity of CP or IP is not indicated, but may be roughly inferred from the inset, which corresponds to the main panel what panel a is to panel b.
  See text for more discussion.}

\end {figure}


In Figures 4a,b, we show line shape simulations for cut b using the Holstein model in the strong-coupling (or the anti-adiabatic) limit where a simple perturbative solution can be obtained \cite {alexandrov}.
For input parameters of the Holstein model, we use $\omega_p$ = 70 meV (66 meV for \eighteeno) and tight binding fit of $\epsilon (k)$ \cite {ek-little-detail} to the \sixteeno~ARPES data.
The interaction parameter $\lambda$ was taken to be isotope dependent \cite {grilli}, 0.9 for \sixteeno~and 0.8 for \eighteeno, in order to give the correct sign for the peak shift.
Note that the peak sharpening for \eighteeno~remains true even if $\lambda$ is taken to be isotope independent.
The essential characteristics of the line shape simulation is the strong multi-phonon ``shake-up'' features, which occur at harmonics of $\omega_p$ (panel a) and which we postulate will broaden, due to phonon continuum and strong electron-electron interaction in solids, into a single peak (panel b), as was also suggested recently \cite {shen}.
In this way, the simulation in panel b successfully reproduces the small IE for sharp low-energy coherent peak (CP) and the large IE for broad high-energy incoherent peak (IP)\@.
As shown, IEs as large as $\approx 30$ meV are generic features of this simulation.
In addition, the magnitude of the ARPES line width is also realistic.
These important results make an explicit demonstration of the multi-phonon physics proposed within a dynamic spin-Peierls picture \cite {gweon-nature}.

The strong coupling theory used here, while capturing the essence of the multi-phonon (polaronic) physics, has two obvious shortcomings in explaining the ARPES kink, a virtually non-dispersive CP with a very small weight $Z \ll 0.1$ and a completely non-dispersive IP \cite {alexandrov}\@.
Both these shortcomings disappear as the interaction strength is reduced \cite {hohenadler,gunnarsson,nagaosa}, and thus it is suggestive that the interaction at optimal doping lie in the intermediate regime, where there is still a significant multi-phonon contribution to the electron self energy {\em and} both CP and IP (especially the latter) show strong dispersions.
While it may be naively expected that the IE will decrease as the interaction strength is reduced, studies \cite {pietronero, mishchenko} show that actually at intermediate couplings the IE is anomalously {\em enhanced}, strengthening our qualitative argument here.

Our results clearly call for a reconsideration of the usual ARPES kink picture based on the MDC analysis within the Migal-Eliashberg theory.
In Figure 4c, we propose a schematic picture that is applicable to intermediate coupling regime as well as Migal-Eliashberg regime, based on numerical simulations \cite {hohenadler, sawatzky, simone}.
The inset shows numerical simulations (Figure 4 of Ref.~\onlinecite {hohenadler}) in the intermediate coupling regime, showing distinct multi-phonon branches reminiscent of multi-phonon peaks in panel a.
Due to the additional broadening mechanisms discussed above, these multi-phonon branches will tend to merge into a single IP branch, as shown in the main panel.
The dispersion of IP at high energy is similar, but not identical, to $\epsilon (k)$ \cite {gunnarsson, mishchenko}.
Within this picture, the kink anomaly still occurs at $\omega_p$, the onset of multi-phonon continuum (gray area), consistent with a small isotope-induced shift \cite {gweon-nature}.
Note that this picture places EDCs as more basic quantities than MDCs.  
Subsequently, the kink energy is best defined as the separation energy between CP and IP branches of EDCs, and the MDC dispersion with a kinky crossover (dotted orange line) is merely a consequence of weight transfer between the two branches.


To conclude, we have shown that IEs in ARPES peak width and peak position give important clues to the nature of electron-lattice interaction in optimally doped \bisco.
Theoretical \cite {fu} and experimental \cite {bianconi-PRL} works further support the emerging picture in which a cooperation of {\em local} lattice distortions and charge/spin order fluctuations produces an interaction effect on electrons beyond the limit of the Migal-Eliashberg theory.

\begin{acknowledgments}


  This work was supported by the Director, Office of Science,
  Office of Basic Energy Sciences, Division of Materials Sciences and
  Engineering of the U.S Department of Energy under Contract No.~DEAC03-76SF00098 and 
  by the National Science Foundation through Grant
  No.~DMR-0349361.

\end{acknowledgments}


\begin{thebibliography}{99}


\bibitem {lanzara} A. Lanzara, et al., Nature {\bf 412}, 510 (2001).
\bibitem {gweon} G.-H. Gweon, et al., J. Phys.~Chem. Solids {\bf 65}, 1397 (2004).
\bibitem {gweon-nature} G.-H. Gweon, et al., Nature {\bf 430}, 187 (2004). 
\bibitem {cuk} T. Cuk, et al., Phys.~Rev.~Lett.~{\bf 93}, 117003 (2004). 
\bibitem {verga} S. Verga, et al., Phys.~Rev.~B {\bf 67}, 054503 (2003).
\bibitem {devereaux} T. P. Devereaux, et al., Phys.~Rev.~Lett.~{\bf 93}, 117004 (2004).
\bibitem {grilli} G. Seibold and M. Grilli, Phys.~Rev.~B {\bf 63}, 224505 (2001).
\bibitem {norman} M. Eschrig and M. R. Norman, Phys.~Rev.~B {\bf 67}, 144503 (2003).
\bibitem {calvani} P. Calvani, et al., Phys.~Rev.~B {\bf 53}, 2756 (1996)
\bibitem {ruani} C. Taliani, et al., Solid State Commun.~{\bf 66}, 487 (1988).
\bibitem {heeger} Y. H. Kim, et al., Phys.~Rev.~B {\bf 36}, 7252 (1987).
\bibitem {bianconi} A. Bianconi and M. Missori, in {\em Phase Separation
in Cuprate Superconductors}, edited by E. Sigmund and A. K. M\"uller
Springer-Verlag, Berlin, (1994), p.~316.
\bibitem {bozovic} I. Bozovic, et al., Phys.~Rev.~Lett.~{\bf 59}, 2219 (1987).
\bibitem {billinge} S. J. L. Billinge and T. Egami, Phys.~Rev.~B {\bf 47}, 14386 (1993); E. S. Bozin, et al., Phys.~Rev.~B {\bf 59}, 4445 (1999).
\bibitem {mihailovic} D. Mihailovic and K. A. M\"uller, {\em High-T$_c$ Superconductivity 1996: Ten Years after the  Discovery}, NATO ASI. Ser. E. Vol. {\bf 343} (Kluwer, Dordrecht, 1997).
\bibitem {imai} T. Imai, et al., Phys.~Rev.~Lett.~{\bf 70}, 1002 (1993).
\bibitem {kochelaev} B. J. Kochelaev, et al., Phys.~Rev.~Lett.~{\bf 79}, 4274 (1997).
\bibitem {shen} K. M. Shen, et al., Phys.~Rev.~Lett.~{\bf 93}, 267002 (2004).
\bibitem {perali} A. Perali, C. Grimaldi, and L. Pietronero, Phys.~Rev.~B {\bf 58}, 5736 (1998).
\bibitem {feng} D. L. Feng, et al., Science {\bf 289}, 277 (2000).
\bibitem {bianconi-PRL} A. Bianconi, et al., Phys.~Rev.~Lett.~{\bf 76}, 3412 (1996).
\bibitem {xingjiang} X.-J. Zhou, et al., Phys.~Rev.~Lett.~{\bf 92}, 187001 (2004).
\bibitem {seidel} A. Seidel, H.-H. Lin, D.-H. Lee, Phys.~Rev.~B {\bf 71}, 220501 (2005).
\bibitem {fu} H. C. Fu, C. Honerkamp, and D.-H. Lee, Europhys.~Lett.~{\bf 75}, 146 (2006).
\bibitem {full-momentum-dep-later} The momentum dependence of the isotope effect beyond the off-nodal cut b is under investigation.
\bibitem {shift} The $^{16}$O data in Figure 1b were shifted by 4 meV in
  energy, in order to offset the superconducting gap difference, with negligible consequence on the
  main reproducible IE at high energy \cite {gweon-nature}.
\bibitem {explain-HWHM-R} We use EDC HWHM$_R$, because the large momentum-asymmetric background intensity at high energy makes unreliable both the MDC width extraction and the EDC left-width.
\bibitem {egami} R. J. McQueeney, et al., Phys.~Rev.~Lett.~{\bf 82}, 628 (1999).
\bibitem {devereaux-private-comm} T. P. Devereaux, private communication.
\bibitem {note-T-dep} The temperature dependence (not shown) of the IE in the
  ARPES width is completely similar to that reported for the ARPES dispersion \cite {gweon-nature}.
\bibitem {fink} A. A. Kordyuk, et al., Phys.~Rev.~B {\bf 71}, 214513 (2005).
\bibitem {mishchenko} A. S. Mishchenko and N. Nagaosa, Phys.~Rev.~B {\bf 73}, 092502 (2006).
\bibitem {fratini} S. Fratini and S. Ciuchi, Phys.~Rev.~B {\bf 72}, 235107 (2005).
\bibitem {alexandrov} A. S. Alexandrov and J. Ranninger, Phys.~Rev.~B {\bf 45}, 13109 (1992).
\bibitem {ek-little-detail} $\epsilon(k)$ was scaled down by a factor of 0.25, to make the strong coupling perturbation theory valid, with little consequence on the calculated line shape.
\bibitem {hohenadler} M. Hohenadler, M. Aichhorn, and W. von der Linden, Phys.~Rev.~B {\bf 68}, 184304 (2003).
\bibitem {gunnarsson} O. R\"osch, and O. Gunnarsson, Eur.~Phys.~Journal B {\bf 43}, 11 (2005).
\bibitem {nagaosa} A. S. Mishchenko and N. Nagaosa, Phys.~Rev.~Lett.~{\bf 93}, 036402 (2004).
\bibitem {pietronero}  P. Paci, et al., Phys.~Rev.~Lett.~{\bf 94}, 036406 (2005).
\bibitem {sawatzky} G. A. Sawatzky, private communication.
\bibitem {simone} S. Fratini, et al., Phys.~Rev.~B {\bf 63}, 153101 (2001).

\end{thebibliography}
\end{document}